\documentclass[useAMS,usenatbib]{mn2e}

\usepackage{amsmath,amssymb}
\usepackage{latexsym}
\usepackage{graphicx}
\usepackage{hyperref}
\def\DPT{{\small DPT}}

\title[]{The effect of deconfinement phase transition on
rotochemical deviations in stars containing mixed phase matter}
\author[Wei Wei Xiao-Ping Zheng]{Wei Wei$^{1,2}$\thanks{E-mail:
weiwei1981@mail.hazu.edu.cn}, Xiao-Ping Zheng$^{1}$\thanks{E-mail: zhxp@phy.ccnu.edu.cn}{}\\
$^{1}$Huazhong Normal University , WuHan ,430079 , China\\
$^{2}$Huazhong Agriculture University , WuHan , 430070 , China}
\begin{document}

\date{}

\maketitle

\label{firstpage}

\begin{abstract}
As a neutron star spins down, its core density increase, changing
the relative equilibrium concentration, and causing deconfinement
phase transition as well. Hadron matter are converted into quark
matter in the interior, which enhances the deviation of chemical
equilibrium state. We study such deviations and its chemical energy
release.Applying to the simulation of cooling neutron stars, we find
the surface effective temperature of neutron stars is promoted
obviously. This implies that the deconfinement phase transition is
able to raise the chemical heating efficiency.
\end{abstract}


\section{Introduction}
Neutron star cooling is an important tool to probe its inner
structure. The comparison of cooling models with observation of
thermal emission provide insight into the EOS of dense matter, and
the signatures of exotic particles\citep{b11}\citep{b16}. A newly
formed neutron star loses the thermal energy through neutrino
emission initially, and is taken over by the surface photon
radiation about $\sim10^{5} $yr after birth. However, heating
sources inside neutron stars can keep neutron stars hot beyond the
standard cooling timescale $\sim10^{7} $yr. Some mechanisms have
been extensively discussed. These include the mutual friction
between superfluid and normal component of the star \citep{b15},
heat released through deconfinement phase transition\citep{b9}and
release of strain energy stored by the solid crust due to spindown
deformation \citep{b3}.

Another of these mechanism is rotochemical heating, firstly proposed
by Reisenegger\citep{b13} and applied to quark stars by
Cheng\citep{b2}. Then Reisenegger improved it, in the framework of
general relativity, by considering the internal structure via
realistic EOSs \citep{b5} and superfluid nucleons core\citep{b12}.
All studies above agree that rotochemical heating is important for
old neutron stars.

Rotochemical heating origins in the deviation from weak reaction
equilibrium due to spin-down compression. As the star spins down,
the decrease of the centrifugal force leads to compression,
increasing the star's internal density. This compression results in
a displacement of the equilibrium concentration of each particle
species and changes the chemical equilibrium state. Reactions that
change the chemical composition drive the system to the new
equilibrium configuration. If the reaction rate is slower than the
change of the equilibrium concentrations due to spin-down
compression, the star is never exactly in chemical equilibrium, so
energy is stored. The excess of energy is dissipated by enhanced
neutrino emission and heat generation.

It has long been hypothesized that some compact stars might actually
be "hybrid stars" containing cores of quark matter\citep{b1}. Unlike
neutron stars with pure hadronic matter and strange stars with pure
strange matter, those hybrid stars probably contain the mixed phase
in light of Glendenning's phase transition theory\citep{b7}. Of
course, the compression of a hybrid star due to spin-down causes
deconfinement phase transition (hereafter \DPT), besides nonequilibium Urca
processes\citep{b14}. The spin-down compression increases the
interior density gradually, which results in the transformation of
hadron matter into quark matter in the interior little by little.
The deconfinement reactions change relative concentrations of the
particles, neutrons, protons, and $u$, $d$, $s$ quarks, and result
in the displacement of equilibrium state as well.
Because neutrons and protons are converted into two flavor quarks,
and it takes a finite amount of time that two flavor quarks  decay
into three. Thus the deconfinement contribution need to be
considered in the calculation of the $\beta$-equilibrium state
deviation.

The purpose of this paper is to study the effect of DPT
on rotochemical heating in mixed phase. In section
2, we attempt a model to describe the deviations from
$\beta$-equilibrium state under the circumstance of DPT.
In section 3, we derive the equations of chemical
and thermal evolution, and discuss an application of a uniform star
model with mixed phase matter. The results and corresponding
explanations are presented in section 4. In section 5, we give our
conclusions and suggest future improvement.

\section{Model}

\begin{figure}
\centering \setlength{\unitlength}{1mm} \leavevmode \hskip  0mm
\includegraphics[width=0.3\textwidth]{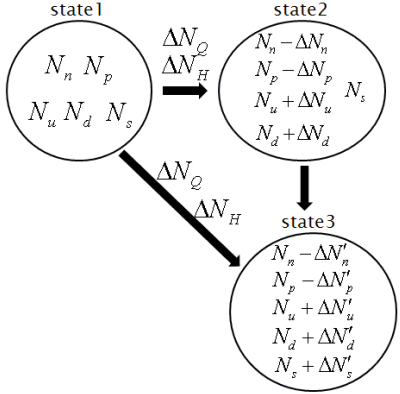}
\caption {The DPT lead to the change of
quark and hadron numbers, which was treated in equilibrium condition
from equilibrium state 1 to equilibrium state 3 in previous study,
$\Delta N_{Q}=\Delta N_{u}'+\Delta N_{d}'+\Delta N_{s}' , \Delta
N_{H}=\Delta N_{n}'+\Delta N_{p}'$. In fact, neutrons and protons
are converted into $u$ and $d$ quarks, without $s$ quarks. So the
deconfinement reactions break the equilibrium state and drive the
system deviating to a nonequilibrium state 2, $\Delta N_{Q}=\Delta
N_{u}+\Delta N_{d}, \Delta N_{H}=\Delta N_{n}+\Delta N_{p}$. Weak
reactions adjust the concentrations of these particles and try to
recover the system to a new equilibrium state 3.}
\label{1fig}
\end{figure}
Hybrid stars have more complicated matter composition than purely
neutron stars, which contains neutrons, protons, $u$, $d$ and $s$
quarks and electrons. The relative concentrations of these particles
are adjusted by the following weak reactions,
\begin{eqnarray}\label{eq:Durca}
u+e^{-}\rightarrow d+\nu_{e} ,\quad
d \rightarrow u+e^{-}+\overline{\nu}_e , \nonumber \\
u+e^{-}\rightarrow s+\nu_{e} ,\quad
s\rightarrow u+e^{-}+\overline{\nu}_e,\\
p + e^{-} \rightarrow n  + \nu_e ,\quad n \rightarrow p + e^{-}
+\overline{\nu}_e\nonumber .
\end{eqnarray}
At the same time, the compression leads to the deconfinement
reactions,
\begin{eqnarray}\label{eq:Decomfinement}
n\rightarrow 2d+u ,\quad p\rightarrow 2u+d .
\end{eqnarray}
Nucleons dissolve into only two flavor quarks, so the relative
concentrations of three flavor quarks need to be adjusted by
non-leptonic reaction,
\begin{eqnarray}\label{eq:Adjust reaction}
u+d\leftrightarrow s+u.
\end{eqnarray}
Thus there are two channels, through which the $\beta$-equilibrium
is violated. One is a direct contribution of spin-down compression
discovered by Reisenegger\citep{b13}. Another is the contribution of
deconfinement reactions caused by the spin-down. The deconfinement
reactions take place firstly and then result in a departure from
$\beta$-equilibrium state.

The deconfinement reactions lead to the change of quark baryon
number density $\Delta n_{q}$ and change of nucleon baryon number
density $\Delta n_{h}$. Since neutrons and protons are converted
into $u$ and $d$ quarks, without $s$ quarks, we have
\begin{eqnarray}\label{eq:density}
\Delta n_{q}=\Delta n_{u}+\Delta n_{d},\quad \Delta n_{h}=\Delta
n_{n}+\Delta n_{p}.
\end{eqnarray}
The reactions(\ref{eq:Decomfinement}) change the concentration of
particles and drive the system deviating from an equilibrium state
(state1) to a nonequilibrium state (state 2). Then, the system tries
to recover to a new equilibrium state (state3) through reactions
(\ref{eq:Durca})and (\ref{eq:Adjust reaction}). We give a
description of these processes in Figure 1. Since the deconfinement
reactions is faster than weak reactions, the system is never exactly
in chemical equilibrium due to deconfinement behavior.

For quarks, the non-leptonic reaction (\ref{eq:Adjust reaction}) and
Durca reactions (\ref{eq:Durca}) are all involved in the evolution
of chemical potential, among which only two processes are needed in
the calculation of the chemical departure, resulting in rotochemical
heating. In the later calculation, we find that the contribution of
the nucleon dominates hybrid star thermal evolution and the chemical
evolution of nucleon is independent of reaction (\ref{eq:Adjust
reaction}). So we don't distinguish the difference between the
non-leptonic reaction and Durca reactions, and choose the Durca
reactions to calculate the chemical potential difference.

To express the chemical deviation quantities, under consideration of
deconfinement reactions, we define the chemical potential of
particle $i$ in nonequilibrium state (state2)as $\mu_{i}^{nq}$ and
$\mu_{i}^{\prime}$ in new equilibrium state (state 3). The chemical
potential differences of three channels in reactions(\ref{eq:Durca})
are:
\begin{eqnarray} \label{eq:potential displacement}
\delta \mu_{D}=\delta\mu_{d}- \delta\mu_{u}- \delta\mu_{e},\nonumber \\
\delta \mu_{S}=\delta\mu_{s}- \delta\mu_{u}- \delta\mu_{e},\nonumber \\
\delta \mu_{N}=\delta\mu_{n}- \delta\mu_{p}- \delta\mu_{e}.\nonumber
\end{eqnarray}
where reaction channels denoted by capitals, particles denoted by
small letters.

We assume $n_{n}=\alpha n_{h}$,$n_{p}=(1-\alpha)n_{h}$ , under beta
equilibrium, then $\Delta n_{n}=\alpha\Delta n_{h}$, $\Delta
n_{p}=(1-\alpha)\Delta n_{h}$, this leads to the following equality,
with constraint by reactions(\ref{eq:Decomfinement}),
\begin{eqnarray} \label{eq:quark fraction}
\Delta n_{d}=\frac{\alpha+1}{2-\alpha}\Delta n_{u},
\end{eqnarray}
Substitute (\ref{eq:quark fraction}) into (\ref{eq:density}), we
obtain
\begin{eqnarray} \label{eq:dengsity change}
\Delta n_{u}=\frac{2-\alpha}{3}\Delta n_{q},\quad \Delta
n_{d}=\frac{1+\alpha}{3}\Delta n_{q}.\nonumber
\end{eqnarray}
At zero temperature, chemical potential of Fermions satisfies
$\mu_{f}^{2}=m_{f}^{2}+(3\pi^{2}n_{f})^{\frac{2}{3}}$, then we
determine chemical potential of particles in each state for density
change and density of a given initial state.\\
state 1:
\begin{eqnarray}
&&\mu_{u}=(3\pi^{2}n_{q})^{\frac{1}{3}}(3+\frac{6\mu_{e}}{\mu_{u}}-\frac{3m_{s}^{2}}{2\mu_{u}^{2}})^{-\frac{1}{3}},\nonumber\\
&&\mu_{d}=\mu_{s}=\mu_{u}+\mu_{e},\nonumber\\
&&\mu_{p}=(3\pi^{2}n_{h})^{\frac{1}{3}}(2+\frac{3\mu_{e}}{\mu_{p}}-\frac{3m_{n}^{2}+3m_{p}^{2}}{2\mu_{p}^{2}})^{-\frac{1}{3}},\nonumber\\
&&\mu_{n}=\mu_{p}+\mu_{e},\nonumber
\end{eqnarray}
state 2:
\begin{eqnarray}
&&\mu_{u}^{nq}=\mu_{u}(1+\frac{2-\alpha}{3}\frac{\Delta n_{q}}{n_q}),\nonumber\\
&&\mu_{d}^{nq}=\mu_{d}(1+\frac{1+\alpha}{3}\frac{\Delta n_{q}}{n_q}),\nonumber\\
&&\mu_{s}^{nq}=\mu_{s},\nonumber\\
&&\mu_{p}^{nq}=\mu_{p}(1+\frac{\mu_{p}^{2}-m_{p}^{2}}{\mu_{p}^{2}}\frac{\Delta n_{p}}{3n_p}),\nonumber\\
&&\mu_{n}^{nq}=\mu_{n}(1+\frac{\mu_{n}^{2}-m_{n}^{2}}{\mu_{n}^{2}}\frac{\Delta n_{n}}{3n_n}),\nonumber
\end{eqnarray}
state 3:
\begin{eqnarray}
&&\mu_{u}^{\prime}=\mu_{u}(1+\frac{\Delta n_{q}}{3n_q}),\nonumber\\
&&\mu_{d}^{\prime}=\mu_{d}(1+\frac{\Delta
n_{q}}{3n_q})-\mu_{e}(1+\frac{\Delta n_{q}}{3n_q})+\mu_{e}'\nonumber
=\mu_{u}^{\prime}+\mu_{e}^{\prime}\nonumber,\\
&&\mu_{s}^{\prime}=\mu_{d}^{\prime}\nonumber,\\
&&\mu_{n}^{\prime}=\mu_{p}^{\prime}+\mu_{e}^{\prime}.\nonumber
\end{eqnarray}
State 2 and 3 can be expressed in light of state 1 and density
change, we immediately write
\begin{eqnarray} \label{eq:potential displacement of particle}
\delta\mu_{u}=\mu_{u}^{nq}- \mu_{u}^{\prime},\nonumber\quad
\delta\mu_{s}=\mu_{s}^{nq}- \mu_{s}^{\prime},\nonumber\quad
\delta\mu_{d}=\mu_{d}^{nq}- \mu_{d}^{\prime},\nonumber\\
\delta\mu_{n}=\mu_{n}^{nq}- \mu_{n}^{\prime},\nonumber\quad
\delta\mu_{p}=\mu_{p}^{nq}- \mu_{p}^{\prime},\nonumber\quad
\delta\mu_{e}=\mu_{e}^{nq}- \mu_{e}^{\prime}.\nonumber
\end{eqnarray}
since state 3 has been defined as equilibrium state, the
$\beta$-equilibrium condition should enter to simplify our
calculations. So
\begin{eqnarray} \label{eq:potential displacement2}
\delta \mu_{D}=\mu_{d}^{nq}- \mu_{u}^{nq}- \mu_{e}^{nq},\nonumber\\
\delta \mu_{S}=\mu_{s}^{nq}- \mu_{u}^{nq}- \mu_{e}^{nq},\nonumber\\
\delta \mu_{N}= \mu_{n}^{nq}- \mu_{p}^{nq}- \mu_{e}^{nq}.\nonumber
\end{eqnarray}

\section{Equations of chemical and thermal evolutions}

According to the discussion above, We first calculate the departure
rate from $\beta $ equilibrium in channel $d\rightarrow
u+e^{-}+\overline{\nu}_{e}$
\begin{eqnarray}
\frac{d\delta\mu_{D}}{dt}=\lim_{\triangle
t\rightarrow0}\frac{1}{\Delta
t}[\mu_{d}^{nq}-\mu_{u}^{nq}-\mu_{e}^{nq}]\nonumber
\end{eqnarray}
To leading order, we obtain
\begin{eqnarray} \label{eq:departure1}
[\frac{d\delta\mu_{D}}{dt}]_{departure}=\frac{2\alpha-1}{3}\frac{\mu_{d}}{n_q}\frac{d
n_q}{d t} +0.\nonumber
\end{eqnarray}
In the similar way, we give
\begin{eqnarray} \label{eq:departure2}
[\frac{d\delta\mu_{S}}{dt}]_{departure}=-\frac{2-\alpha}{3}\frac{\mu_{s}}{n_q}\frac{d
n_q}{d t}.\nonumber
\end{eqnarray}
\begin{align} \label{eq:departure3}
[\frac{d\delta\mu_{N}}{dt}]_{departure}=&[\frac{(3\pi^2n_n)^\frac{2}{3}}{((3\pi^2n_n)^\frac{2}{3}
+m_n^2)^\frac{1}{2}}\nonumber\\
&-\frac{(3\pi^2n_p)^\frac{2}{3}}{((3\pi^2n_p)^\frac{2}{3}+m_p^2)^\frac{1}{2}}
-(3\pi^2n_p)^{\frac{1}{3}}]\frac{1}{3n_h}\frac{dn_h}{dt}.\nonumber
\end{align}
On the other hand, the departure is preferred to recover its
equilibrium through reactions (\ref{eq:Durca}). Therefore, we have
\begin{eqnarray} \label{eq:restore1}
[\frac{d\delta\mu_{d}}{dt}]_{restore}=-\Gamma_{d}E_{xx}^{d}(\mu),\nonumber
\end{eqnarray}
\begin{eqnarray} \label{eq:restore2}
[\frac{d\delta\mu_{s}}{dt}]_{restore}=\Gamma_{s}E_{xx}^{s}(\mu),\nonumber
\end{eqnarray}
\begin{eqnarray} \label{eq:restore3}
[\frac{d\delta\mu_{n}}{dt}]_{restore}=-\Gamma_{n}E_{xx}^{n}(\mu),\nonumber
\end{eqnarray}
where $x$ indicates $d$, $s$, $n$ fraction of baryon density, $
E_{xx}$ is partial derivative of chemical energy per
baryon\citep{b2}. In the hadron-quark mixed phase, hadron and quark
undergo direct Urca processes\citep{b9}. $\Gamma_{d} $,
$\Gamma_{s}$and $\Gamma_{n}$ are the reaction rates per
baryon\citep{b4}\citep{b8}. Combine departure and corresponding
restore, we immediately get the time evolution equations of the
chemical potential differences.
\begin{eqnarray}
\frac{d\delta\mu_{D}}{dt}=\frac{2\alpha-1}{3}\frac{\mu_{d}}{n_q}\frac{dn_q}{dt}-\Gamma_{d}
E_{xx}^{d}(\mu),\label{eq:deviation equation1}
\end{eqnarray}
\begin{eqnarray}
\frac{d\delta\mu_{S}}{dt}=-\frac{2-\alpha}{3}\frac{\mu_{s}}{n_q}\frac{dn_q}{dt}+\Gamma_{s}
E_{xx}^{s}(\mu),\label{eq:deviation equation2}
\end{eqnarray}
\begin{eqnarray}\label{eq:deviation equation3}
\begin{aligned}
\frac{d\delta\mu_{N}}{dt}=&[\frac{(3\pi^2n_n)^\frac{2}{3}}{((3\pi^2n_n)^\frac{2}{3}+m_n^2)^\frac{1}{2}}-\frac{(3\pi^2n_p)^\frac{2}{3}}{((3\pi^2n_p)^\frac{2}{3}+m_p^2)^\frac{1}{2}}\\
&-(3\pi^2n_p)^{\frac{1}{3}}]\frac{1}{3n_h}\frac{dn_h}{dt}-\Gamma_{n}E_{xx}^{n}(\mu),
\end{aligned}
\end{eqnarray}
Equation (\ref{eq:deviation equation1})and (\ref{eq:deviation
equation2}) show that $D$ and $S$ channels obey their evolutional
equations respectively, differing from Cheng $\&$ Dai \citep{b2}
discussion of strange quark matter. Inclusion of deconfinement
effect causes such serious difference. To solve these equations,
thermal equation is necessary, reads
\begin{eqnarray}\label{eq:thermal evolution}
\begin{aligned}
c_{v}\frac{dT}{dt}=(\Gamma_{d}\delta\mu_{D}+&\Gamma_{s}\delta\mu_{S}+\Gamma_{n}\delta\mu_{N})\\
&-(\epsilon_{d}+\epsilon_{s}+\epsilon_{n})-\dot{E}_{\gamma},
\end{aligned}
\end{eqnarray}
where the three terms represent the chemical energy released by the
reactions, the energy radiated by neutrino and antineutrino, and the
energy in photons from the surface of the star. $c_{v}$ is the
specific heat of dense matter $c_{v}=\chi c_{q}+(1-\chi)c_{h}$. Now
we need to treat $\mu_{f}$,$ n_{q}$,$ n_{h}$ as the function of
total baryon number density. In 1992, Glendenning first presented
the existence of first -order phase transition with Gibbs
construction. He applied the relativistic mean field theory for
hadronic matter and MIT for quark matter to construct the phase
diagram. Baryon densities of each phase in mixed phase are depicted
in his figure7 \citep{b6}. We here consider a uniform star model for
simulation and take baryon density $n=1.0 fm^{-3}$.We approximate,
around the baryon density, $n_{q}\approx0.3431+0.8228n$,
$n_{h}\approx0.07143+0.7428n$,
$\mu_{d}\approx\frac{1}{3}\mu_{n}\approx\frac{1}{3}(4.826+1.44n)$.\\
From Reisenegger\citep{b13}, the total baryon number density changes
with the spin-down by
\begin{eqnarray}
\frac{1}{n}\frac{dn}{dt}\approx-\frac{\Omega}{G\rho_{c}}\frac{d\Omega}{dt}.\nonumber
\end{eqnarray}
where $\rho_{c}$ is central density of the stars.

\section{Results}

\begin{figure}
\setlength{\unitlength}{1mm} \leavevmode \hskip  0mm\centering
\includegraphics[width=0.5\textwidth]{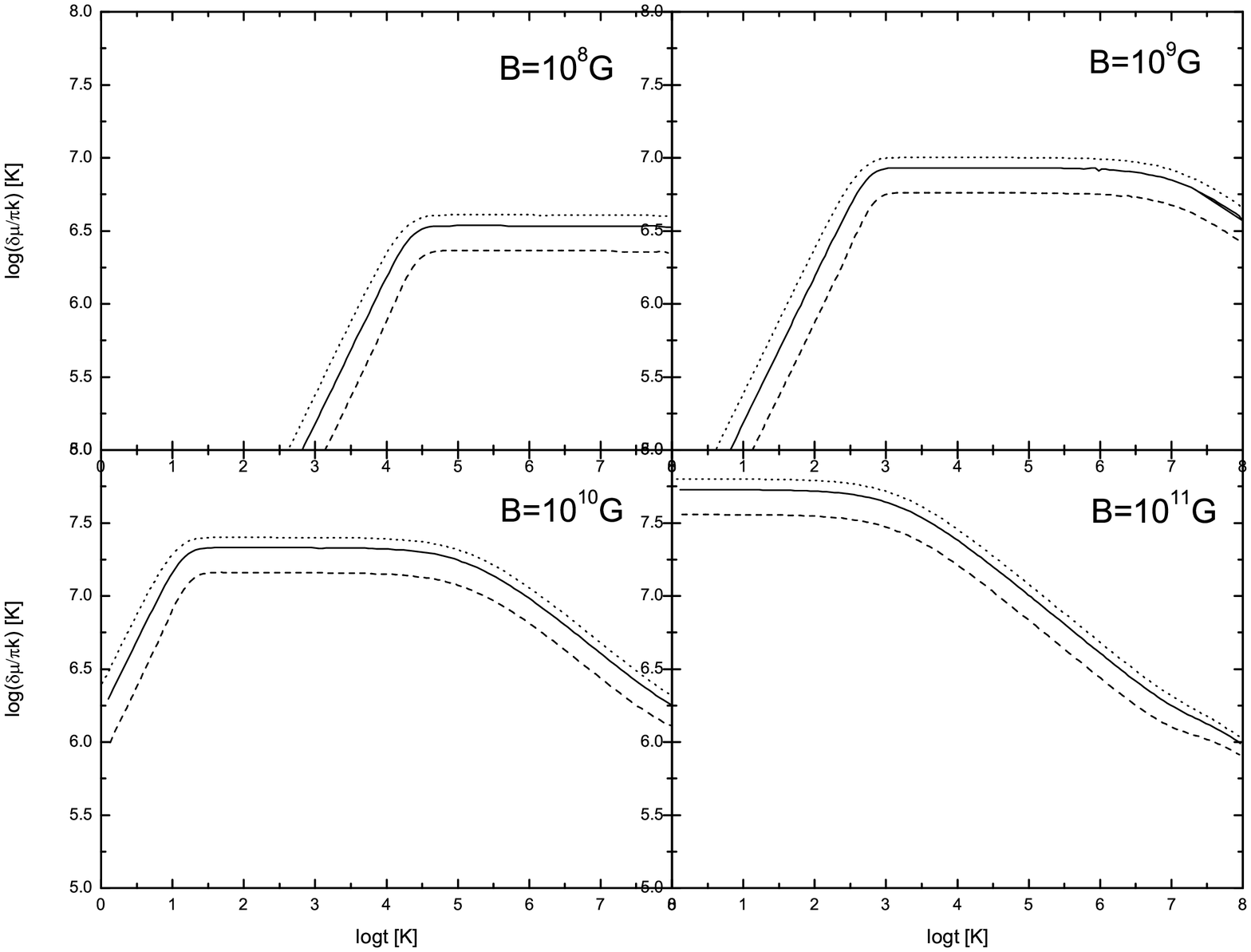}
\caption{Chemical evolution. The variables $\delta\mu/\pi k$(a
measure for the depature from chemical equilibrium) is plotted
logarithmically as functions of time.The solid line, dashed line and
dotted line are chemical evolution for $S$ channel, $D$ channel and
nucleon channel respectively.The four panels correspond to different
magnetic field strengths: (a)$B=10^{8} G$; (b) $B=10^{9} G$;
(c)$B=10^{10} G$; (d) $B=10^{11} G$.}
\label{2fig}
\end{figure}

We give an initial rotation period $Pi=1 ms$ and
$\alpha=\frac{8}{9}$, the latter corresponding to the threshold for
direct Urca process to be allowed\citep{b10}. After giving the
strong coupling constant, $ s$-quark mass, and the surface magnetic
field strength , we integrate the coupling evolution equations
(\ref{eq:deviation equation1})$\thicksim$(\ref{eq:thermal
evolution}).

Figure 2 shows the chemical evolution in hadron-quark mixed phase.
The solid line, dashed line and dotted line are chemical evolution
for $S$ channel, $D$ channel and nucleon channel, respectively.
Since neutrons and protons are converted into $u$ and $d$ quarks,
without $s$ quarks, the chemical potential departures of $S$ channel
is higher than that of $D$ channel. The four panels correspond to
four different magnetic field strengths:(a)$B=10^{8} G$, (b)
$B=10^{9} G$, (c)$B=10^{10} G$, (d) $B=10^{11} G$. The stronger the
magnetic field strength, the more the kinetic energy of rotation
converting into chemical energy ,the larger the chemical potential
departure, the earlier the appearance of heating effect. The
platform indicates a relative quasi-equilibrium process. And the
rapid dropping of chemical departure weaken the rotochemical heating
at later times for strong magnetic. But rotochemical heating is
still substantial for $B=10^{8} G$ at later times. Figure 3 gives
the chemical potential departure of nucleon channel with magnetic
field strengths $B=10^{8} G$ and $B=10^{10} G$, to compare
deconfinement and spin-down situations. Obviously, the deconfinement
reactions enlarge the chemical potential departure.
\begin{figure}
\setlength{\unitlength}{1mm}
\leavevmode\centering
\hskip  0mm
\includegraphics[width=0.5\textwidth]{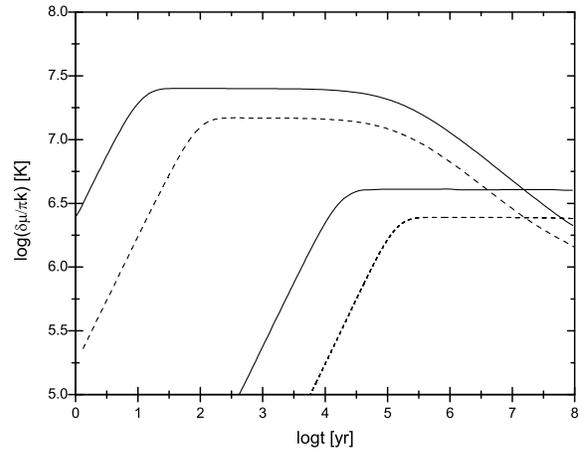}
\caption
{Chemical potential departure is plotted logarithmically as a function of time for considering
DPT (\textit{solid line} )and without it(\textit{dashed line} ),
with magnetic field strengths $B=10^{8} G$ and $B=10^{10} G$.}
\label{3fig}
\end{figure}
\begin{figure}
\setlength{\unitlength}{1mm}
\leavevmode\centering
\hskip  0mm
\includegraphics[width=0.5\textwidth]{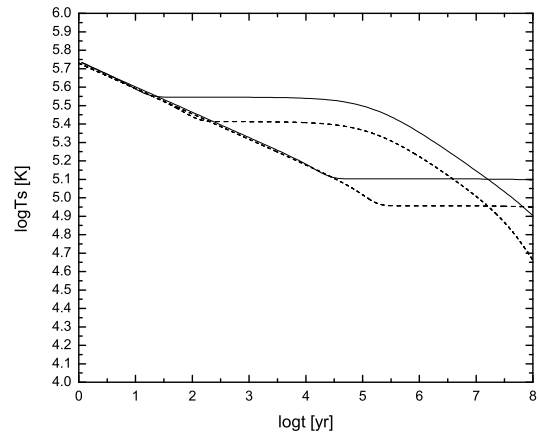}
\caption
{Effective surface temperature as a function of time, for rotochemical heating considering
DPT (\textit{solid line} )and traditional rotochemical heating
(\textit{dashed line} ),with magnetic field strengths $B=10^{8} G$ and $B=10^{10} G$.}
\label{4fig}
\end{figure}

The chemical evolution influence the cooling of neutron stars, when
the chemical heating mechanism is considered. The DPT enhance the
chemical potential departure of particles, and effect the thermal
evolution of stars containing mixed phase matter as well. Figure 4
shows the evolution of the effective surface temperature for uniform
hybrid star with rotochemical heating under consideration of DPT,
and the traditional case. Although we use a toy model in which
several approximations are made, the effective surface temperature
of the star is about 40 percent higher than the traditional one, and
the heating effect appears earlier.
\begin{figure}
\setlength{\unitlength}{1mm}
\leavevmode\centering
\hskip  0mm
\includegraphics[width=0.5\textwidth]{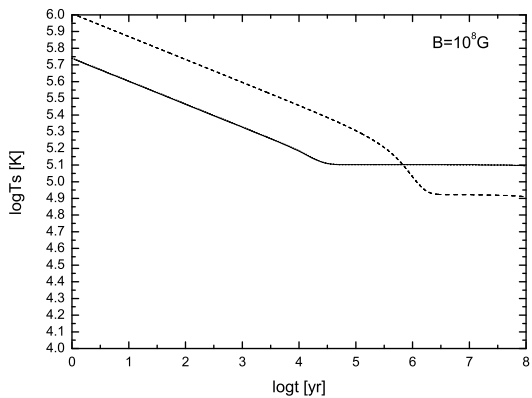}
\caption {Effective surface temperature as a function of time, with
magnetic field strengths $B=10^{8} G$, for only considering quark's
chemical deviation and emission(\textit{dashed line} ),only
considering nucleon's(\textit{dotted line} ) and considering
both(\textit{solid line} ).}
\label{5fig}
\end{figure}

We also analysis the contributions of quarks and nucleons by
comparing temperature magnitudes. Figure 5 gives the results with
$B=10^{8} G$. Since nucleon's curve is close to full one, the
contribution of the nucleon dominates hybrid star thermal evolution.

Figure 6 shows the evolution of the effective surface temperature
with different magnetic field strengths $B=10^{8}G$(\textit{dashed
line}), $10^{9}G$(\textit{dotted line}),
$10^{10}G$(\textit{dashed-dotted line}),
$10^{11}G$(\textit{dashed-dotted-dotted line}), and in the absence
of chemical heating. One sees that the effect of rotochemical
heating with DPT can be substantial at later times for stars with weak fields
and still important at earlier times for stars with strong fields.
\begin{figure}
\setlength{\unitlength}{1mm}
\leavevmode\centering
\hskip  0mm
\includegraphics[width=0.5\textwidth]{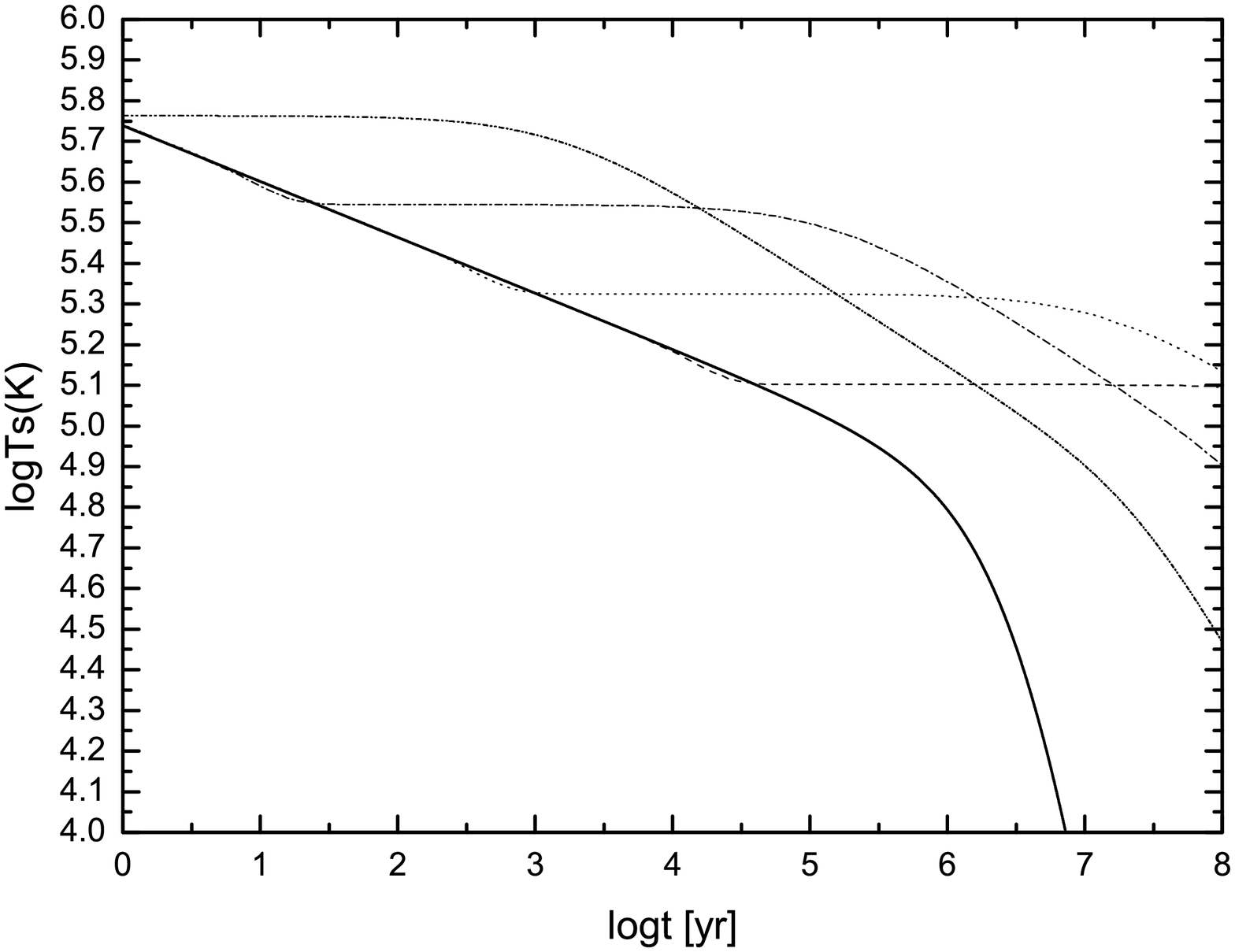}
\caption {Effective surface temperature as a function of time, with
no heating (solid lines) or rotochemical heating with magnetic field
strengths $B=10^{8} G(\textit{dashed line} )$, $10^{9}
G(\textit{dotted line} )$, $10^{10} G(\textit{dashed-dotted line}
)$, and $10^{11} G(\textit{dashed-dotted-dotted line} )$. }
\label{6fig}
\end{figure}

\section{Conclusions}

Besides the spin-down, an adjoint of DPT
indeed causes the departure from $\beta$-equilibrium in mixed phase
matter. This mechanism is investigated in this paper. The new
mechanism makes a substantial promotion in effective surface
temperature. We usually believe hybrid stars cool down rapidly
following standard cooling theory. Our results show that a hybrid
star should has much higher temperature when the chemical heating with DPT is
considered. Although we consider a uniform density model to
facilitate our understanding of the mechanism, the results are very
close to the hybrid stars containing large mixed phase core if their
mean density equals the assumed uniform density.

In our calculations, we set $\alpha$ as a constant and only consider
the contribution of the DPT in chemical
deviation. The new mechanism makes a greater contribution than the
spin-down one. So the direct contribution of spin-down compression
is not important for rotochemical deviations in stars containing
mixed phase matter and can be ignored in calculations.

This paper focus on studying a new mechanism in heating generation
inside hybrid stars, our consideration of initial period $Pi=1 ms$
is reasonable for millisecond pulsars, but is too short for
classical pulsars with higher magnetic field strengths. So the more
reasonable initial period is needed in the accurate calculations
comparing with the observations.

Accurate calculations are worth being expected. Our future works
would study the thermal evolution of hybrid star including DPT,
taking the structure of star into account in the frame of general
relativity, using realistic EOSs of dense matter. The results thus
would compare with the observations.\\

\section*{Acknowledgments}
The authors are grateful to K.S.Cheng for valuable discussions.
This work is supported by the National Natural Science Foundation of
China under Grant Nos.11079008. Wei Wei also thanks the support of
NSFC under Grant Nos.11003005.

\bsp

\label{lastpage}


\begin{thebibliography}{99}
\bibitem[\protect\citeauthoryear{Baym \& Chin}{1976}]{b1} Baym G., Chin S.A., 1976, Phys.Lett.B, 62, 241
\bibitem[\protect\citeauthoryear{Cheng \& Dai}{1996}]{b2} Cheng,K.S., Dai,Z.G., 1996, ApJ, 468, 819
\bibitem[\protect\citeauthoryear{Cheng et al.}{1992}]{b3} Cheng,K.S., Chau,W.Y., Zhang,J.L., Chau,H.F., 1992, ApJ, 396, 135
\bibitem[\protect\citeauthoryear{Dai et al.}{1993}]{b4} Dai,Z.G., Lu,T., Peng,Q.H. 1993, Phys.Lett.B, 319, 199
\bibitem[\protect\citeauthoryear{Fernandez \& Reisengger}{2005}]{b5} Fernandez,R., Reisenegger,A., 2005, ApJ, 625, 291
\bibitem[\protect\citeauthoryear{Glendenning}{1992}]{b6} Glendenning,N.K., 1992, PRD, 46,
1247
\bibitem[\protect\citeauthoryear{Glendenning}{2000}]{b7} Glendenning,N.K., 2000, Compact Stars,Nuclear Physics,Particle Physics,and General Relativity. New York
\bibitem[\protect\citeauthoryear{Haensel}{1992}]{b8} Haensel P., 1992, A\&A, 262, 131
\bibitem[\protect\citeauthoryear{Kang \& Zheng}{2007}]{b9} Kang,Miao, Zheng,Xiao-Ping, 2007, MNRAS, 375, 1503
\bibitem[\protect\citeauthoryear{Lattimer et al.}{1991}]{b10} Lattimer,J.M., Pethick,C.J., Prakash,M., Haensel,P., 1991, Phys.Rev.Lett., 66, 2701
\bibitem[\protect\citeauthoryear{Page et al.}{2006}]{b11} Page D., Geppert U., Weber F., 2006, Nucl. Phys. A, 777, 497
\bibitem[\protect\citeauthoryear{Petrovich \& Reisengger}{2010}]{b12} Petrovich,C., Reisenegger,A., 2010, A\&A, 521,
77
\bibitem[\protect\citeauthoryear{Reisenegger}{1995}]{b13} Reisenegger,A., 1995, ApJ, 442, 749
\bibitem[\protect\citeauthoryear{Stejner et al.}{2009}]{b14} Stejner M., Weber F., Madsen J., 2009, ApJ, 694, 1019
\bibitem[\protect\citeauthoryear{Shibazaki \& Lamb}{1989}]{b15} Shibazaki,N., Lamb,F.K., 1989, ApJ, 346, 808
\bibitem[\protect\citeauthoryear{Yakovlev \& Pethick}{2004}]{b16} Yakovlev,D.G., Pethick,C.J., 2004, ARA\&A, 42, 169
\end{thebibliography}
\end{document}